\documentclass[prc,twocolumn,showpacs,nofootinbib,aps,longbibliography,superscriptaddress]{revtex4-2}
\usepackage[utf8]{inputenc}
\usepackage[T1]{fontenc}
\usepackage[normalem]{ulem}
\usepackage{dcolumn}
\usepackage{bm}
\usepackage{graphicx}
\usepackage{epstopdf}
\usepackage{wrapfig}
\usepackage{array} 
\usepackage{listings}
\usepackage[para,online,flushleft]{threeparttablex}
\usepackage{booktabs,dcolumn}
\usepackage{dsfont}

\usepackage{textpos}
\usepackage{booktabs}
\usepackage{multirow,bigdelim}
\usepackage{newtxtext,newtxmath,amsmath}

\usepackage{xcolor}
\definecolor{pastelgray}{rgb}{0.81, 0.81, 0.77}
\definecolor{beaublue}{rgb}{0.9, 0.9, 0.93}
\usepackage{orcidlink}

\usepackage{hyperref}

\hypersetup{colorlinks=true, citecolor=blue, linkcolor=black, urlcolor=blue}

\usepackage{multibib}
\newcites{methods}{References for Methods}

\usepackage{tikz}
\usetikzlibrary{matrix,positioning,calc}

\def\spacingset#1{\renewcommand{\baselinestretch}%
	{#1}\small\normalsize} \spacingset{1}

\usepackage{url}

\newcommand{\PRLsep}{\noindent\makebox[\linewidth]{\resizebox{0.3333\linewidth}{1pt}{$\bullet$}}\bigskip}
\usepackage{bm} 

\newcommand{\cb}{\bm{c}}

\newcommand{\Sb}{\bm{S}}

\newcommand{\Ub}{\bm{U}}
\newcommand{\Vb}{\bm{V}}    

\newcommand{\xb}{\bm{x}}    
\newcommand{\yb}{\bm{y}}    

\newcommand{\omegab}{\bm{\omega}} 

\newcommand{\mean}{\phi_0}
\newcommand{\meanelements}{\mean(x_i)}
\newcommand{\Xelements}{X_{i,k}^0}
\newcommand{\Xcelements}{X_{i,k}^c}
\newcommand{\Xb}{\bm{X^0}}
\newcommand{\Xbc}{\bm{X^c}}

\newcommand{\PCParams}{b}

\newcommand{\LDMparams}{\boldsymbol{a}}

\newcommand{\ModelTotal}{m}
\newcommand{\PCTotal}{p}

\begin{document}

\title{Model orthogonalization and Bayesian forecast mixing via Principal Component Analysis}





\author{P. Giuliani}

\email{giulianp@frib.msu.edu}
\affiliation{Facility for Rare Isotope Beams, Michigan State University, East Lansing, Michigan 48824, USA}

\author{K. Godbey}
\email{godbey@frib.msu.edu}
\affiliation{Facility for Rare Isotope Beams, Michigan State University, East Lansing, Michigan 48824, USA}

\author{V. Kejzlar}

\affiliation{Mathematics and Statistics Department, Skidmore College, Saratoga Springs, New York 12866, USA}

\author{W. Nazarewicz}
\email{witek@frib.msu.edu}
\affiliation{Facility for Rare Isotope Beams, Michigan State University, East Lansing, Michigan 48824, USA}
\affiliation{Department of Physics and Astronomy, Michigan State University, East Lansing, Michigan 48824, USA}

\begin{abstract}
One can improve predictability in the unknown domain by combining forecasts of imperfect complex computational models using a Bayesian statistical machine learning framework. In many cases, however, the models used in the mixing process are similar. In addition to contaminating the model space, the existence of such similar, or even redundant, models during the multimodeling process can result in misinterpretation of results and deterioration of predictive performance. In this work we describe a method based on the Principal Component Analysis that eliminates model redundancy. We show that by adding model orthogonalization to the proposed Bayesian Model Combination framework, one can arrive at better prediction accuracy and reach excellent uncertainty quantification performance.

\end{abstract}

\flushbottom
\maketitle

\section{Introduction}
Modeling is a crucial part of many scientific disciplines.
Within the  framework of the scientific method,
models are designed to create postdictions about  past data, to describe phenomena,  and make predictions about the future observations.
In many cases, several alternative (and competing) models  are available to describe a given physical phenomenon. These models might be based on different theoretical foundations, calibrated to different datasets, involve different computational algorithms, and often will have a different accuracy when it comes to forecasting (i.e., postdictions or predictions).

Choosing one of the models either arbitrarily or using off-the-shelf model selection method  leads to poor uncertainty quantification (UQ). To this end, combining together a set of different models is advisable
~\cite{Hoge2019,StackingGelman18,BANDmanifesto}. One of the key aspects of multimodeling is the choice of individual models whose forecasts are combined and the elimination of very similar, or even redundant, models is a challenge.

The objective of this work is to find the effective number of models in the model set and determine their relative contributions to the combined forecast obtained within the Bayesian setting.
To this end,  we use Principal Component Analysis (PCA). We shall demonstrate that incorporating model preselection and model orthogonalization via PCA into the multi-model framework leads to:
(i) faster and scalable forecasting (only the reduced set of orthogonalized models is mixed); (ii) improved computational robustness of multi-modeling; (iii) increased interpretability through elimination of similar models; (iv) improved predictive performance as properly orthogonalized models are less prone to overfitting.

Below, we first briefly review the fundamentals of our multi-model framework. As an illustration, we apply our approach to a pedagogic example of predicting nuclear binding energies using a simple analytic model and study common modeling scenarios. Finally, we show the opportunities provided by our method for practitioners on a case study of predicting binding energies using a set of realistic models based on the nuclear density functional theory (DFT).

\section{Multimodeling}


\subsection{Reasonable models}

Let us consider a set of models $\mathcal{M}_1, \dots, \mathcal{M}_\ModelTotal$
which are used to forecast observations of a physical process at locations $x_i\in \mathcal{X} \subset \mathbb{R}^n$, $i=1,\ldots,n$, where 
$\mathcal{X}$ indicates the input domain of observations. As the main goal of  this study is to develop a framework for quantified extrapolations, we introduce a notion of
 ``reasonable models'', i.e., the models which: (i)  are well suited to provide  sound quantified  forecasts  within the input domain 
$\mathcal{X}_0 \subset \mathcal{X} $
in which experimental observations exist, and  (ii) have sound physical/microscopic foundations enabling  extrapolations and UQ outside of $\mathcal{X}_0 $ into the unknown domain $\mathcal{X}^*=\mathcal{X}  - \mathcal{X}_0$.
Each model $\mathcal{M}_k$ is calibrated to a dataset within the input subdomain $\mathcal{X}_k  \subset \mathcal{X}_0$.
The condition (ii) excludes phenomenological, many-parameter formulae fitted to experimental data, which yield uncontrolled extrapolations
in the unknown domain $\mathcal{X}^*$.

\subsection{Combining forecasts}
In this section, we briefly overview three basic approaches to combining forecasts of reasonable models (for a comprehensive discussion, the reader is referred to Refs.~\cite{Hoge2019,StackingGelman18}).
In general, the goal of forecast combination is to use several models to predict observations $y(x)$ of a physical process at new locations $x^*\in \mathcal{X}^*$
using information from both  measurements/observations $\yb = [y(x_1), \dots, y(x_{n})]$ and model calculations at the input locations $x_i\in \mathcal{X}_0$.

One common approach for combining forecasts is \textit{Bayesian model averaging} (BMA)
~\cite{Hoeting1999,Wasserman2000,Fragoso2018}.  
Here, the resulting prediction is given by a mixture of individual models' posterior predictive distributions where the BMA model weights reflect the fit of a statistical model to data independently of the set of available models and are obtained by marginalizing over model parameters (i.e., using Bayesian evidence). However, BMA relies on theoretical assumptions which are inappropriate for approximate modeling of complex systems (e.g., one of the candidate models is a ``true'' model that perfectly describes the physical reality). 

The \textit{Bayesian model mixing}  (BMM) framework, an extension of Bayesian stacking ~\cite{LeClarke2017, StackingGelman18,BANDmanifesto, StackingGelman22},  implicitly assumes that while none of the models ${\cal M}_k$ is true, the underlying physical process is well captured by a linear combination of the models.
A resulting
statistical model can be written as 
~\cite{Fernandez2002,StackingGelman22,BANDmanifesto,Kejzlar2023}:
\begin{equation}
    \label{eqn:LMM-model}
    y(x_i) = \sum_{k=1}^\ModelTotal \omega_k(x_i)
    f_k(x_i) + \sigma_i \epsilon_i,
\end{equation}
where $\sigma_i$ represents the scale of the error of the mixture model (possibly including experimental and theoretical errors), $\epsilon_i \overset{\mathrm{iid}}{\sim} N(0,1) $, $f_1(x_i), \dots, f_\ModelTotal(x_i)$ are  forecasts for the datum $y(x_i)$ provided by the $\ModelTotal$ theoretical models considered, and $\omegab(x_i) \equiv [\omega_1(x_i), \dots, \omega_\ModelTotal(x_i)]$ are their respective {\it weights} which are often adjusted to fulfill the simplex constraint:
 \begin{equation}\label{simplex}
\omega_1, ..., \omega_\ModelTotal \geq 0,~~ \sum_{k=1}^\ModelTotal  \omega_k = 1.
\end{equation}

The weights can, in principle, depend locally on the input domain or can be global, i.e., domain-independent.
The distribution of BMM model weights additionally depends on the modeling choice for $\omegab$  and the set of models considered ~\cite{Kejzlar2023}. 
As demonstrated in previous studies ~\cite{Bates1969,Minka2002,Kejzlar2023}, combining models using BMM  outperforms BMA in terms of both prediction accuracy and UQ. Consequently, in this work's case studies
 we do not pursue the  BMA strategy.

The \textit{Bayesian model combination} (BMC) strategy
aims to find a combined forecast that outperforms individual forecasts by hoping that systematic deficiencies of different models will compensate. Here, the focus is on the  overall performance rather than the relation of the models ${\cal M}_k$ to the true model. 
In BMC, one assumes the mixture model
in the form ~\cite{Granger1984,Hoge2019}:
\begin{equation}
    \label{eqn:LMC-model}
    y(x_i) = \sum_{k=1}^\ModelTotal c_k(x_i)
    f_k(x_i) + f_0 + \sigma_i\epsilon_i,
\end{equation}
where $\cb=[c_k]$ are model {\it amplitudes} and $f_0$ is an optional constant term.
If $f_0=0$, BMC is reduced to the unrestricted combined forecast whose amplitudes $\cb$ can be determined by unrestricted chi-square minimization or by constructing a Bayesian posterior distribution given the data. In general, the amplitudes of BMC do not have to be positive
~\cite{Bates1969,Granger1984}.
Some  applications of BMC~\cite{Monteith2011} impose the simplex constraint (\ref{simplex}) ; this results in a worsened performance ~\cite{Granger1984}. 

\subsection{Model similarity and redundancy}

In many cases, physics models may have a similar mathematical foundation but their parameters are calibrated using different methodologies. It is also possible that  models are in fact identical in spite of their different formulation. Consider, e.g., (i) a model given by a polynomial of order $n$ (Taylor expansion) and (ii) a model given by a Legendre multipole expansion of order $n$. Both models are manifestly identical, if calibrated to the same dataset. This extreme situation is referred to as \textit{model redundancy}~\cite{Burnham2002}.

In addition to ``polluting'' the model space $[{\cal M}_k]$,  the existence of redundant or similar models during the multimodeling process  
can result in difficulties with obtaining reliable inferences and hence misinterpretation of results and deterioration of predictive performance. The standard application of BMA will particularly suffer from this situation given that each model weight is calculated independently of the set of available models, allowing for overemphasis of model classes with repeated representations.
Consequently, adding model preselection and orthogonalization to model combination pipelines is important and, as we show in the following, relatively straightforward.

\section{Model orthogonalization}\label{Sec: orthogonalization}

Principal Component Analysis (PCA) ~\cite{jolliffe2002principal} and Singular Value Decomposition (SVD)~\cite{blum_hopcroft_kannan_2020} are two related methods that have become essential tools for data compression,  signal processing, data visualization, feature selection, and dimensionality reduction across science and engineering~\cite{brunton2019data}. 
In the past \cite{Clyde1996}, PCA has been specifically applied to model orthogonalization; see also other PCA applications to combining forecasts \cite{Poncela2011,Stock2012}, including a recent application to nuclear mass models~\cite{Wu2024}.

For the purpose of this work, we will use PCA to identify the first $\PCTotal$ principal components, or directions of maximal variability, across a set of $\ModelTotal$ theoretical mass models ($\PCTotal\leq \ModelTotal$). 
We first consider forecasts of the $\ModelTotal$ different models: $f_k(x_i)$ ($i=1, \dots, n)$, where $n$ is the number of model results. For our specific application, these forecasts consist of the $n\sim 600$  computed nuclear binding energies of different even-even nuclei characterized by the number of protons $Z$  and neutrons $N$, i.e., the domain of interest is defined by $x_i= ({Z}_i, {N}_i)$, see Ref.\cite{Neufcourt18-1}. We arrange these model results into a matrix $\Xb= (\Xelements)$, where a fixed column represents the forecast of a single model $f_k$ across all $n$ measurements, while a fixed row represents the predictions of all $\ModelTotal$ models on a fixed mass
 of nucleus $x_i$.
From this matrix, we construct the centered matrix $\Xbc \equiv (\Xcelements)$,
\begin{equation}\label{eq:centering}
     \Xcelements =  \Xelements - \meanelements,
\end{equation}
by subtracting the average 
$\mean(x)$ of all the models (columns),
\begin{equation}\label{eq:mean}
\meanelements=\frac{1}{\ModelTotal}\sum_{k=1}^\ModelTotal \Xelements =\frac{1}{\ModelTotal}\sum_{k=1}^\ModelTotal f_k(x_i).
\end{equation}
The vector $\mean(\boldsymbol{x})$,
where  $\boldsymbol{x} =[x_i]$ denotes the list of inputs, represents the average forecast of all models and in principle contains the main features that a reasonable physical model should have. The deviations from this average, contained in the matrix $\Xbc$, serve to characterize  individual  models. 
The matrix $\Xbc$ can be expressed in the singular value decomposition form:
\begin{align}\label{Eq: SVD}
    \Xbc_{{n\times \ModelTotal}} &= \Ub_{{n\times n}} \ \  \Sb_{{n \times \ModelTotal}} \ \  \Vb^T_{{\ModelTotal \times \ModelTotal}} \notag \\
    &\approx \ \ \hat \Xbc_{{n\times \PCTotal}}  = \hat \Ub_{{n\times  \PCTotal}} \ \  \hat \Sb_{{ \PCTotal \times  \PCTotal}} \ \  \hat \Vb^T_{{ \PCTotal \times  \ModelTotal}}.
\end{align}
In Eq.~(\ref{Eq: SVD}), 
the reduced-dimension matrix  $\hat \Xbc$  optimally~\cite{eckart1936approximation,brunton2019data} approximates the original matrix $\Xbc$ by keeping only the first $\PCTotal\leq \ModelTotal$ singular values $s_j$ of $\Sb$. The total number $\PCTotal$ of components  kept can be chosen in several ways (see for example the partial sum criterion~\cite{jolliffe2002principal}), and in this study we treat it as a hyperparameter that is selected by analyzing the performance of the overall model across a validation set.

\begin{figure}[htb] 
	\begin{centering}
	\includegraphics[width=0.4\textwidth]{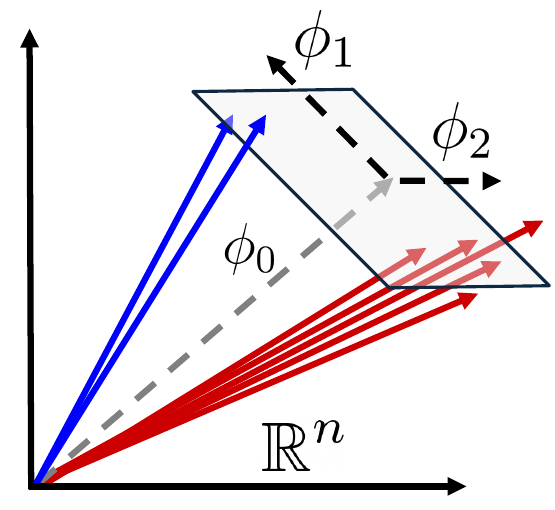}
		\caption{Schematic representation of the PCA approach for the model combination. Here, two model classes consists of 2 and 5 models, respectively, and are represented as vectors in a  space $\mathbb{R}^n$. This collection of 7 models is approximated in the affine space (grey rectangle) spanned by the constant $\phi_0$ term (dashed light-grey arrow) and the two principal components $\phi_1$ and $\phi_2$ (dashed black arrows). 
  }
		\label{fig: FigurePCA}
	\end{centering}
\end{figure}

Once the truncation is done, the retained $\PCTotal$ principal components   are obtained by the columns of $\hat \Ub$. 
We label these components as $\phi_j(x)$ (with $j=1, \dots, \PCTotal$). 
As illustrated in in Fig.~\ref{fig: FigurePCA}, the forecast of any of the original models can be  approximated by a linear combination of this smaller set of $\PCTotal$ orthogonal components $\phi_j(\boldsymbol{x})$ identified by the SVD algorithm, plus the original average forecast:
\begin{equation}\label{eq:PC expansion}
    f_k(\boldsymbol{x}) \approx \mean(\boldsymbol{x}) + \sum_{j=1}^\PCTotal \nu^{(k)}_j \phi_j(\boldsymbol{x}), \ \ \text{for } k=1, \dots, \ModelTotal.
\end{equation}
The coefficients $\nu^{(k)}_j$ can be obtained by multiplying the k$^{\text{th}}$ column of $\hat\Vb^T$ by the respective singular values of $\hat\Sb$. 
This reduction presents several advantages for BMM and BMC that we will discuss in the following.

\subsection{Global model mixing and model combination with principal components}

Conveniently, we can construct a combined model $f^\dagger$ by mixing (globally) the identified principal components instead of the original forecasts:
\begin{equation}
    \label{eqn: BMM PC}
    f^\dagger(\xb;\boldsymbol{\PCParams}) = \mean(\xb) + \sum_{j=1}^\PCTotal \PCParams_j
    \phi_j(\xb),
\end{equation}
where $\PCParams_j$ corresponds to the global weight of the $j^{\text{th}}$ principal component $\phi_j$.
Note that since each principal component $\phi_j$ is itself a linear combination of the original forecasts $f_k$, we can 
express Eq.~\eqref{eqn: BMM PC} in a BMM-like form:
\begin{equation}
    \label{eqn:LMM-model-reverse}
     f^\dagger(\xb;\boldsymbol{\PCParams}) = \sum_{k=1}^\ModelTotal \omega_k(\boldsymbol{\PCParams})f_k(\xb),
\end{equation}
where the  $\ModelTotal$ weights $\omega_k$ depend on the $\PCTotal$ latent variables $\boldsymbol{\PCParams} =(\PCParams_1, \dots, \PCParams_\PCTotal)$ as degrees of freedom:
\begin{equation}
    \boldsymbol{\omega}(\boldsymbol{b}) = \frac{1}{m}\mathbf{1} + \Big(\hat \Vb_{{\ModelTotal \times  \PCTotal   }} \hat \Sb^{-1}_{{ \PCTotal \times  \PCTotal}} \Big) \boldsymbol{b},
\end{equation}
where $\mathbf{1}$  is the all-ones
vector of dimension $m$ with all elements
 equal to 1.

By construction, the sum of the weights $\omega_k$ in Eq.~\eqref{eqn:LMM-model-reverse} adds up to 1, i.e., it  satisfies  the second part of the simplex constraint~\eqref{simplex}. Indeed, since $\phi_0$ is the average over all models, it contributes with $m\times \frac{1}{m}=1$ to the total sum of the weights, while every principal component $\phi_j$ is itself a linear combination of the columns of the matrix $\Xbc$, each of which adds net zero total sum of model weights, see Fig.~\ref{fig: FigurePCA}. 
We note, however,  that while the weights $\omega_k$
fulfill the second simplex constraint, they do not have to be positive.

Given the available data, the weights $\boldsymbol{\PCParams}$ 
can be jointly estimated ~\cite{BDA}, with an assumed combined model error scale $\sigma$, within a Bayesian framework:
\begin{equation}\label{Eq: posterior}
    p(\boldsymbol{\PCParams}, \sigma|\boldsymbol{y}) \propto p(\boldsymbol{y}|\boldsymbol{\PCParams}, \sigma)p(\boldsymbol{\PCParams}, \sigma).
\end{equation}
Here, $p(\boldsymbol{y}|\boldsymbol{\PCParams}, \sigma)$ is the data likelihood function with the standard Gaussian-noise assumption as in Eq.~\eqref{eqn:LMM-model}, namely
\begin{align}
\label{eqn:Likelihood}
p(\boldsymbol{y}|\boldsymbol{\PCParams}, \sigma) &\propto \sigma^{-n} \exp\left(-\frac{1}{2}\chi^2\right), \notag \\
\chi^2 &= \sum_{i=1}^n \frac{(f^\dagger(x_i;\boldsymbol{\PCParams}) - y(x_i))^2}{\sigma^2},
\end{align}
and $p(\boldsymbol{\PCParams}, \sigma)$ is the joint prior distribution of the mixing weights and the common error scale $\sigma$. 

The assumption that the deviations between our model predictions $ f^\dagger(x_i;\boldsymbol{\PCParams})$ and the observations $y(x_i)$ follow independent Gaussian distributions with the same noise scale ($\sigma_i\equiv\sigma$) is a common choice for calibrating nuclear models with extremely precise data such as nuclear masses~\cite{Kejzlar2020,Mcdonnell2015,Giuliani2023}. This assumption can be easily modified for different applications of the framework if needed.

A reasonable choice for $p(\boldsymbol{\PCParams}, \sigma)$ is multivariate Gaussian prior distribution for $\boldsymbol{\PCParams}$, informed by the empirical distribution of the original weights $\nu_j^{(k)}$ when reproducing the models in~\eqref{eq:PC expansion}, and a Gamma prior distribution for the precision $1/ \sigma^2$ parametrized by a shape parameter $\nu_0$ and a scale parameter $\sigma_0$, see Ref.~\cite{hoff2009first}. We create a weakly informed prior by selecting $\nu_0=10$ and $\sigma_0=2$\,MeV
for the first case study, and  $\nu_0=10$ and $\sigma_0=0.5$ MeV
for the realistic case.
These choices do not appreciably impact the obtained posterior distributions. Notably, our procedure allows for an efficient approximation of the posterior distribution~\eqref{Eq: posterior} using the standard Gibbs sampler, because the multivariate Gaussian and Gamma priors $p(\boldsymbol{\PCParams}, \sigma)$ together with the likelihood~\eqref{eqn:Likelihood} form the classical normal-inverse-gamma semiconjugate model (see Chapter 9 of~\cite{hoff2009first}).

This prior choice effectively resembles BMC in the context of the combined model~\eqref{eqn:LMM-model-reverse}, because the weights $\omega_k(\boldsymbol{\PCParams})$ are unrestricted in the sense that they can be positive and negative. If one wishes to  impose a further constraint on the model weights, such as the simplex constraint~\eqref{simplex}, the original multivariate Gaussian prior distribution for $\boldsymbol{\PCParams}$ can be easily modified so that it is zero whenever the constraint is not satisfied.

The combined model~\eqref{eqn: BMM PC}, referred to as BMC+PCA in the following, parameterizes an approximated manifold of the currently developed theoretical models through the mixing weights $\boldsymbol{\PCParams}$, as illustrated by the grey affine subspace in Fig.~\ref{fig: FigurePCA}. In the Bayesian context, this combined model, equipped with a posterior probability distribution for the weights~\eqref{Eq: posterior}, has the appealing statistical interpretation of representing an overarching distribution of plausible models that can explain the observed data. Within this framework, each original forecast  $f_k$ could be interpreted as a random -- not necessarily independent -- draw from this distribution, and through Eq.~\eqref{Eq: posterior}  we aim to create a more informed quantified prediction for new observables. Furthermore, performing the mixing on the principal components instead of the original forecasts has the advantage of both filtering out parametric directions that could be associated with noise instead of important features, and prevents the combined-model parameters $\boldsymbol{\PCParams}$ of becoming ill-conditioned in the presence of similar or redundant forecasts $f_k$. We demonstrate these features in the next section.
 
\section{Results}\label{sec: results}


\subsection{Case Study I: Redundant and similar models} \label{Sec: LDM}

To test the proposed global BMC framework~\eqref{eqn: BMM PC}, we first consider nuclear binding energy forecasts generated  by several variants of the analytic Liquid Drop Model (LDM) ~\cite{Weizsacker1935,Ring1980}.
Within this 7-parameter model, the
binding energy of a nucleus, ${\cal E}(N,Z)$, is given by ~\cite{Reinhard2006}:
\begin{align}\label{eqn:ldm_extended_c}
    {\cal E}(N,Z;\LDMparams) &= a_{\text{vol}}A + a_{\text{surf}}  A^{2/3} + a_{\text{curv}}  A^{1/3} + a_{\text{sym}}  I^2A \notag \\
    &\quad + a_{\text{ssym}} I^2  A^{2/3} + a_{\text{sym}}^{(2)}I^4 A + a_{\rm Coul} \frac{Z^2}{A^{1/3}},
\end{align}
where
   $A = N+Z$ is the mass number and  $I = (N-Z)/A$ is the isospin excess.
The parameters $\LDMparams=$  [$a_{\text{vol}}$, $a_{\text{surf}}$, $a_{\text{curv}}$,  $a_{\text{sym}}$,  $a_{\text{ssym}}$, $a_{\text{sym}}$, $a_{\rm Coul}$] have a well-defined physical meaning; they represent volume,  surface, curvature, symmetry, surface-symmetry, second-order-symmetry, and Coulomb  terms respectively. 

\begin{figure*}[htb] 
	\begin{centering}		\includegraphics[width=1\textwidth]{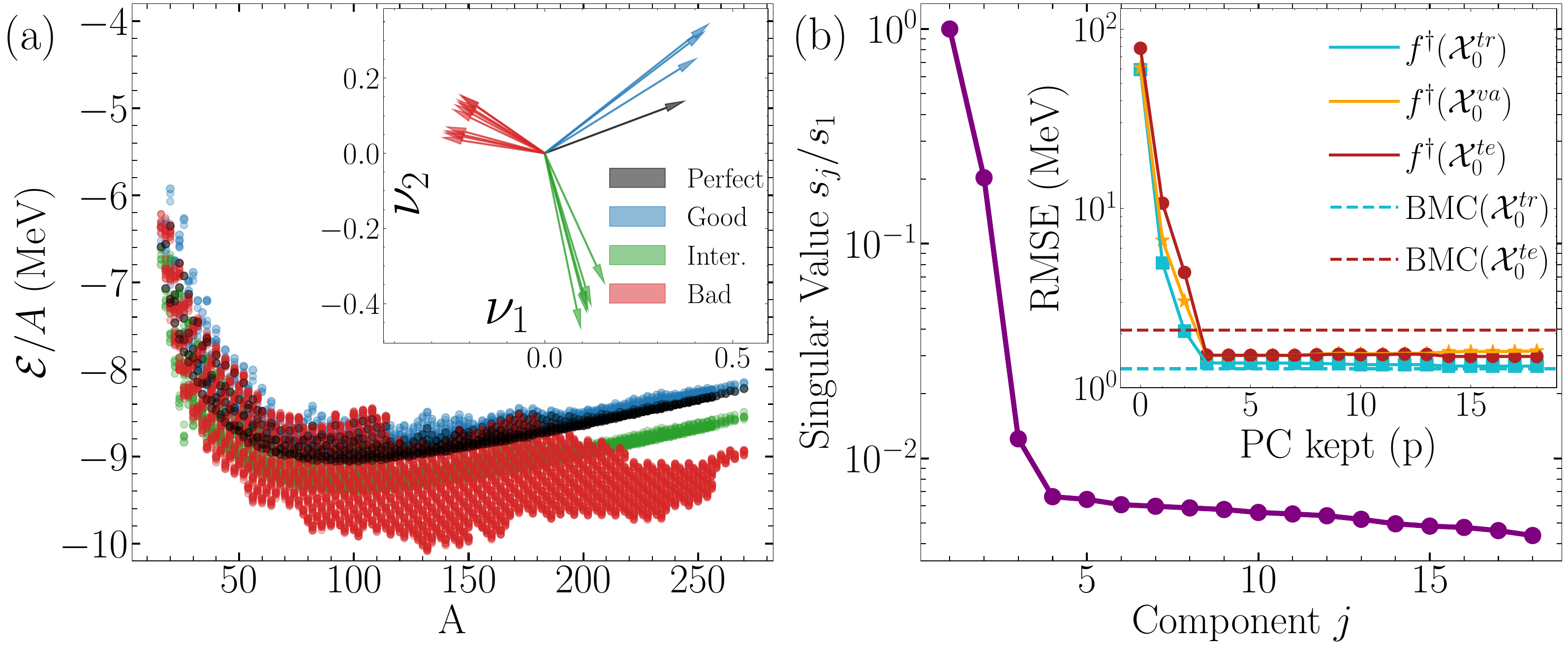}
		\caption{Illustration of  S3 of Table~\ref{tab: Scenarios}. 
  {\bf (a)}: Forecasts of the binding energy per nucleon produced by  19 different models: 1 Perfect model (black),  3 Good models (blue), 5 Intermediate models (green), and 10 Bad models (red). The spread of the results comes from the noise terms added.
  The inset  shows the projection $\nu_j^{(k)}$ defined in 
 Eq.~\eqref{eq:PC expansion} for each of the 19 models  onto the first two principal components, clearly identifying the existence of three model  classes, with the perfect model and three good models being nearly aligned. {\bf (b)}: Decay of the singular values $s_j$.
  The inset  shows the evolution of the RMSE~\eqref{eq: RMSE} for the training (cyan blue squares), validation (yellow stars), and testing (dark red circles) datasets as  the number of principal components  kept in the expansion~\eqref{eqn: BMM PC} is increased (0 corresponds to $\phi_0$). The BMC+PCA results are marked by solid lines.
  The dashed lines show the RMSE obtained when combining all 19 models without projecting on principal components (pure BMC), which  shows signs of overfitting: lower RMSE for the training dataset while higher RMSE for the testing set.
  }
		\label{fig: LDM-PC}
	\end{centering}
\end{figure*}

To test various common modeling scenarios, we create four model classes of the form:
\begin{equation}\label{eq: model maker}
     y_{\text{th}}^{(t)}(N,Z) =  {\cal E}(N,Z;\LDMparams^{(t)}), \ \ \text{for }t\in \{\text{P, 
     G, I, B} \},
\end{equation}
where the model-class index  P, G, I, and B, stands  for ``Perfect", ``Good", "Intermediate", and ``Bad" models, respectively. These labels reflect how close these models are to the the reference  model that generates the synthetic data.
Each model class has parameters centered around parameters $\LDMparams^{(t)}$ defined in Table~\ref{tab: Scenarios}. For some scenarios, to obtain non-degenerate forecasts, the parameters
are shifted by a small random amount $\delta \LDMparams^{(t)}$, which is a Gaussian with  a width of  2\textperthousand\ of  $\LDMparams^{(t)}$. In some cases, 
 we also add a  Gaussian noise term with the width  $\sigma_{\rm noise} = 1$\,MeV. These two  sources of error: the shift in the parameters and the overall Gaussian noise, simulate a situation in which models within the same class yield predictions that deviate both in a coherent  ($\delta \LDMparams^{(t)}$) and uncorrelated  ($\sigma_{\rm noise}$) way. (The spread of model predictions due to these sources of noise is shown in Fig.~\ref{fig: LDM-PC}(a).)
The reference forecast
$y_{\rm true}(N,Z)$ consists of  $n=629$ binding energies of even-even nuclei with $8\leq Z \leq 102$  computed with  the SkO parametrization with the noise term with $\sigma_{\rm noise}$ added.

\begin{table}[htb]
\caption{Four model classes used in this study and their respective parameters  as defined in Table~I of Ref. ~\cite{Reinhard2006}. The  models belonging to the class ``Bad"  are based on the NL1 parametrization with the terms $\{ a_{\text{sym}},a_{\text{ssym}},  a^{(2)}_{\text{sym}}\}$ set to zero;  we label this parametrization  as NL1$^\star$. Three scenarios (S1-S3) considered are shown on
columns 3-5 that list the number 
$N_{\rm rep,k}$ 
of repeated (redundant) models belonging to  class $k$; the number of principal components $p$ kept; and the individual model weights $\omega_k$~\eqref{eqn:LMM-model-reverse} obtained by maximizing the likelihood~\eqref{eqn:Likelihood} in the case without noise  (both $\delta \LDMparams^{(t)}$ and $\sigma_0$ are set to zero). Since no noise was added, every repeated model within the same class has identical weight. For each scenario,
$\sum \omega_k N_{\rm rep,k}=1$. The weights from  S1 are correctly re-distributed among the repetitions in S2, i.e., $(\omega_k N_{\rm rep,k})_{S1}=(\omega_k N_{\rm rep,k})_{S2}$ for each $k$.
The perfect model is selected in S3.
}\label{tab: Scenarios}
\begin{ruledtabular}
\begin{tabular}{c|c|cc|cc|cc}
\begin{tabular}[c]{@{}c@{}}Model\\ class \end{tabular} & \begin{tabular}[c]{@{}c@{}}Parameter\\ center $\LDMparams^{(t)}$\end{tabular} & \multicolumn{2}{c|}{\begin{tabular}[c]{@{}c@{}}S1\\ $\PCTotal=2$\end{tabular}} & \multicolumn{2}{c|}{\begin{tabular}[c]{@{}c@{}} S2\\ $\PCTotal=2$\end{tabular}} & \multicolumn{2}{c}{\begin{tabular}[c]{@{}c@{}}S3\\ $\PCTotal=3$\end{tabular}} \\ \cline{3-8} 
            &                                                                           & $N_{\rm rep}$  & $\omega_k$ & $N_{\rm rep}$  & $\omega_k$ & $N_{\rm rep}$  & $\omega_k$ \\ \hline
Perfect           & SkO                                                                       & 0        & -         & 0        & -         & 1        & 1.000         \\
Good           & SLy4                                                                      & 1        & 0.710         & 3        & 0.237         & 3        & 0.000         \\
Intermediate           & NL1                                                                       & 1        & 0.309          & 5        & 0.062         & 5        & 0.000         \\
Bad           & NL1$^\star$                                                            & 1        & $-$0.019         & 10       &  $-$0.002         & 10       & 0.000         \\
\end{tabular}
\end{ruledtabular}
\end{table}

We study three scenarios S1-S3, which are described in Table~\ref{tab: Scenarios}
and illustrated in Fig.~\ref{fig: LDM-PC} for scenario S3. In these scenarios, we use different numbers of models in each class with the objective of demonstrating that the proposed algorithm works as intended. The singular values of the SVD~\eqref{Eq: SVD} can give an initial estimate of the expected number of effective components, as shown Fig.~\ref{fig: LDM-PC}(b). The projections of each model on the identified principal components $\phi_k$ can visually help identify model classes, as is also shown in the inset of Fig.~\ref{fig: LDM-PC}(a).

For the remaining part of this section, we use the BMC+PCA model  \eqref{eqn: BMM PC}. 
The  synthetic data are separated into three groups: training dataset ($\mathcal{X}_0^{tr}$) with 300 datapoints, validation dataset ($\mathcal{X}_0^{va}$) with 71 datapoints, and testing dataset ($\mathcal{X}_0^{te}$) with 258 datapoints, as is often done in  machine learning applications~\cite{brunton2019data}, including studies focusing on nuclear mass models~\cite{Neufcourt18-2,Neufcourt2020b,Kejzlar2023,Saito2024}.
The specific way the three sets for our work was chosen (see Panel (a) of Figure~\ref{fig:4panel}) reflects that one of the main objectives of our model forecast combination lies in model extrapolation into the region in which experimental information does not exist. By dividing the sets in this way, we provide a more stringent test of how the combined model's performance will evolve as we go further away from the region where data currently exists.

To select the number of principal components kept $\PCTotal$, we study how it  impacts the root mean squared error (RMSE):
\begin{equation}\label{eq: RMSE}
    \text{RMSE} = \sqrt{\frac{1}{n}\sum_{i=1}^n \Big(f^\dagger(x_i;\boldsymbol{\PCParams}_0) -y_{\rm true}(x_i) \Big)^2}.
\end{equation}

The parameters $\boldsymbol{\PCParams}_0$ are chosen to maximize the likelihood function~\eqref{eqn:Likelihood}  on the training dataset $\mathcal{X}_0^{tr}$. The RMSE is then computed across all the sets and the number of components $\PCTotal$ is chosen based on the performance on the validation set. The inset 
in Fig.~\ref{fig: LDM-PC}(b) shows RMSE($\PCTotal$) for  S3.  The RMSE computed for the three datasets  saturate after $\PCTotal=3$, as expected. Indeed,  there are  only four distinct model classes in S3. The training RMSE is always lower, since the parameters $\boldsymbol{\PCParams}_0$ are fitted to it, and the validation RMSE correctly serves as a proxy for the expected RMSE in the extrapolated testing dataset.

\subsection{Case Study II: Realistic nuclear mass models}

We now turn to a set of realistic  models of nuclear binding energy.
For this study, we have chosen 15 realistic computational  models that represent a few classes of theoretical frameworks that see broad use. The models are specified in  Table~\ref{tab: RMSE Realistic}, with their respective RMSE for each of the three datasets.
As the model orthogonalization and mixing strategies only require precomputed data across a range of nuclei, we pull forecasts  directly from published datasets.
The specific mass models chosen and their parameters are those from 
the MassExplorer database ~\cite{massexplorer} and HFB-24~\cite{HFB24dataset} and
FRDM-12  ~\cite{FRDM2012} mass tables. 

\begin{figure*}[htb!] 
	\begin{centering}
\includegraphics[width=1\textwidth]{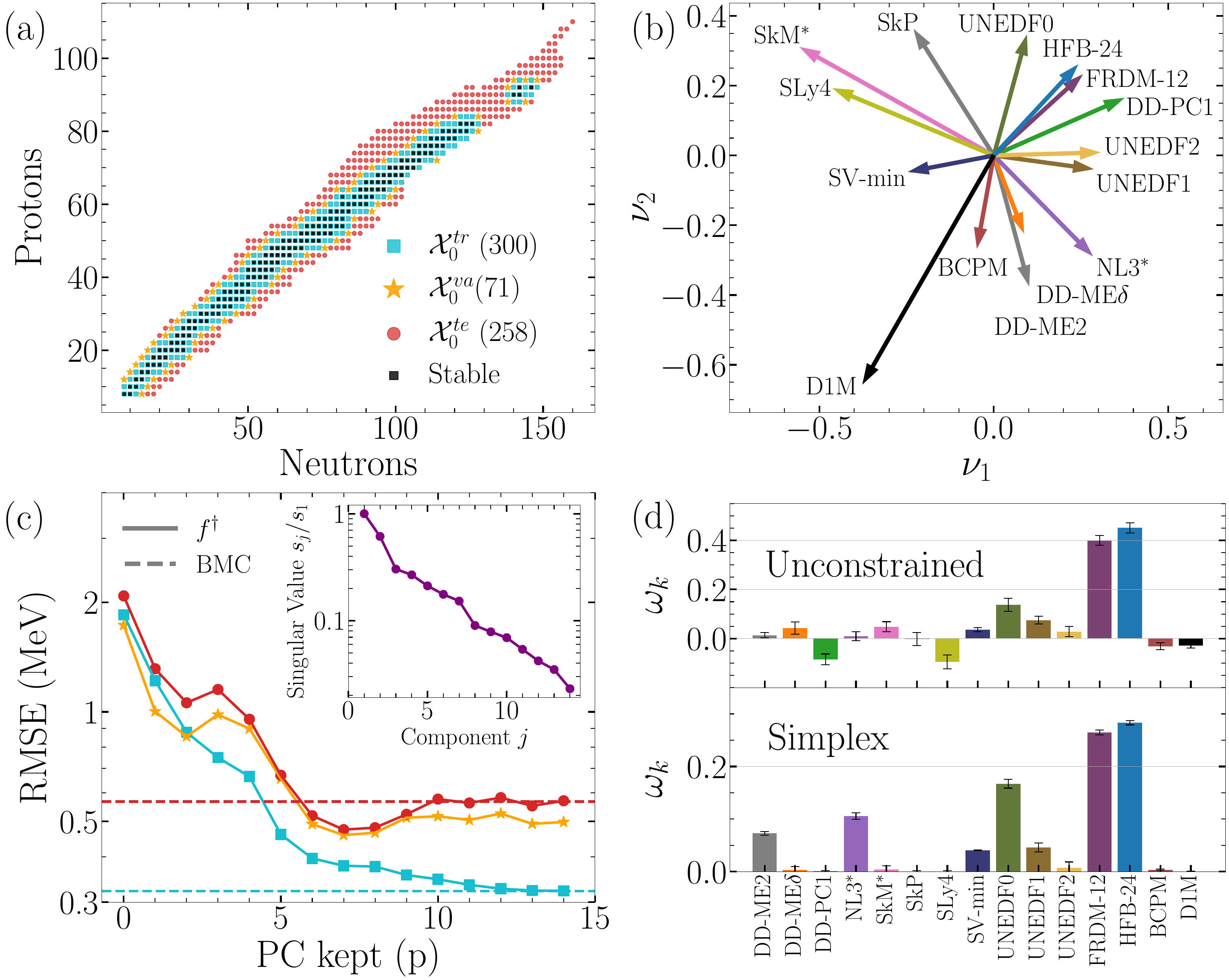}
		\caption{Case study II results. 
{\bf (a)}: training (squares), validation (stars), and testing (circles) datasets 
   of binding energies of
    629  even-even nuclei used in this study.
The stable isotopes are marked by small black squares. 
{\bf (b)}: Projections $\nu_1$ and $\nu_2$ of  15 realistic models of the nuclear  binding energy into the first two principal components. This representation allows us to visualize inter-model relationships. 
{\bf (c)}: Similar to Fig.~\ref{fig: LDM-PC}(b) but for the realistic mass models. The colors and symbols follow the same convention as in panel (a), with solid lines representing the  BMC+PCA model of Eq.~\eqref{eqn: BMM PC} and dashed lines representing the BMC of Eq.~\eqref{eqn:LMC-model} with $f_0=0$.
{\bf (d)}: Distribution of the weights $\omega_k$ for the individual models in the expansion~\eqref{eqn:LMM-model-reverse} in the unconstrained (top) and simplex-constrained (bottom) settings (see Eq.~\eqref{simplex}). The vertical error bars represent a 95$\%$ region obtained from the sampled posterior.
        }
		\label{fig:4panel}
	\end{centering}
\end{figure*}

The domains of the experimental datasets used are shown in Fig.~\ref{fig:4panel}(a). As in the case study I, we divide the 629 data points  into training, validation, and testing sets. We perform the SVD on the forecasts produced by the set of 15 models restricted to the training set and show their projections $\nu_1$ and $\nu_2$ on the first 
two principal components in Fig.~\ref{fig:4panel}(b). The  nearly-exponential decay of the singular values~\eqref{Eq: SVD} is shown in  the inset of panel (c). Note that the SVD is performed after the centering procedure, thus the principal components  in panel (b) should be interpreted as the variability about the average in the space of predictions. While it is tempting to draw conclusions on model similarity from these projections, it can only be said that the forecasts themselves are similar if two vectors are close. The models HFB-24 and FRDM-12, for instance, share little in common when it comes to the underlying form and theoretical assumptions, though their predictive capability is very similar due to their parametric expressivity.
The UNEDF series of interactions (UNEDF0, UNEDF1, UNEDF2) is another interesting case study in that they all have a very similar functional form, but are based on different calibrations. This difference directly manifests in the projections where UNEDF0 is nearly orthogonal to the other two models in the same family despite their close functional relationship. Additional information could likely be gleaned from a more targeted study of certain nuclei, but the global dataset of nuclear binding energies  does not immediately reveal model specific physical insights.

We now proceed to create the combined BMC+PCA model $f^\dagger$ as specified in Eq.~\eqref{eqn: BMM PC}. To this end, we analyze the RMSE performance by maximizing the likelihood~\eqref{eqn:Likelihood} of the combined model as we vary the number of principal components kept. 
Based on the result shown in Fig.~\ref{fig:4panel}(c), we retain  $\PCTotal=7$  principal components for the combined model for the rest of this analysis.  Indeed, as the number of components is increased beyond  7, the validation and test errors grow, suggesting overfitting. The dashed lines show the RMSE obtained when combining all 15 models without projecting on principal components, which also shows signs of overfitting: a lower RMSE for the training dataset and a higher RMSE for the test dataset. It is to be noted that  even though the singular values shown in the inset of panel (c) decay nearly exponentially, they do not experience a rapid drop as in Fig.~\ref{fig: LDM-PC}(b). In this case, the RMSE($p$)  behavior for  the validation set provides a good metric to determine the number of principal components that are needed to optimally model the  experimental values outside of the training set. 

Figure~\ref{fig:4panel}(d) shows the weights $\omega_k$ of each model for both the unconstrained and simplex-constrained case. The vertical error bars show the $95\%$ credible intervals obtained from sampling the weights using Eq.~\eqref{Eq: posterior}. 
In the unconstrained case, several models dominate the combination, though with significant cancellation of the model amplitudes.
In the case of the simplex constrained combination, we see a similar behavior for the models with the largest weights, FRDM-12 and HFB-24 that make up roughly 50\% of the combined model.
The starkest difference is in the effective nullification of many of the other models across the model space, leaving only a few active 
(in our case: 7) in the combination.
The specifics of  weight distributions  naturally depend on the number of principal components that are retained, though the general behavior seems to be consistent across different active subspaces.

\begin{table}[ht]
\caption{Model performance quantified by the RMSE (in MeV). Columns 2, 3, 4, and 5 show the RMSE across the training, validation, testing, and full dataset, respectively. The two bottom rows show the performance of the combined BMC+PCA model $f^\dagger$, in the unconstrained and simplex variant. These RMSE were calculated by averaging the model predictions from the visited posteriors in Eq.~\eqref{Eq: posterior}, and then using Eq.~\eqref{eq: RMSE}.}
\begin{ruledtabular}
\begin{tabular}{@{}l|cccc@{}}  
Model       & RMSE $\mathcal{X}_0^{tr}$ & RMSE $\mathcal{X}_0^{va}$ & RMSE $\mathcal{X}_0^{te}$ & RMSE $\mathcal{X}_0$ \\ 
 \hline \\[-6pt]
DD-ME2~\cite{dd-me2}    & 2.47       & 2.48            & 2.25      & 2.38     \\
DD-ME$\delta$~\cite{dd-medelta} & 2.46       & 2.19            & 2.18      & 2.32     \\
DD-PC1~\cite{dd-pc1}    & 1.94       & 1.77            & 2.19      & 2.03     \\
NL3$^*$~\cite{nl3s}   & 2.18       & 2.15            & 3.59      & 2.84     \\
SkM$^*$~\cite{SKMstar}   & 4.91       & 6.34            & 9.75      & 7.42     \\
SkP~\cite{SKP}    & 3.06       & 3.50            & 4.41      & 3.72     \\
SLy4~\cite{SLY4}   & 4.53       & 4.84            & 6.27      & 5.34      \\
SV-min~\cite{SVMIN}     & 2.97       & 2.99            & 3.98      & 3.42   \\
UNEDF0~\cite{UNEDF0}  & 1.47       & 2.02            & 1.51      & 1.56    \\
UNEDF1~\cite{UNEDF1} & 1.82       & 1.83            & 2.06      & 1.92   \\
UNEDF2~\cite{UNEDF2} & 1.83       & 1.66            & 2.04      & 1.90  \\
FRDM-12~\cite{FRDM2012} & 0.62       & 0.62            & 0.65      & 0.63   \\
HFB-24~\cite{HFB24dataset}  & 0.52       & 0.51            & 0.52      & 0.52  \\
BCPM~\cite{bcpm}   & 2.57       & 2.34            & 2.44      & 2.49    \\
D1M~\cite{d1m}    & 5.02       & 4.91            & 5.63      & 5.27    \\
 \hline \\[-6pt]
$f^\dagger [\PCTotal=7]$ & 0.38      & 0.46            & 0.47      & 0.43        \\
$f^\dagger \text{(simplex)} [\PCTotal=7]$ & 0.74      & 0.69            & 0.71      & 0.72      
\end{tabular}
\end{ruledtabular}
\label{tab: RMSE Realistic}
\end{table}

\begin{figure*}[htb] 
	\begin{centering}
	\includegraphics[width=1\textwidth]{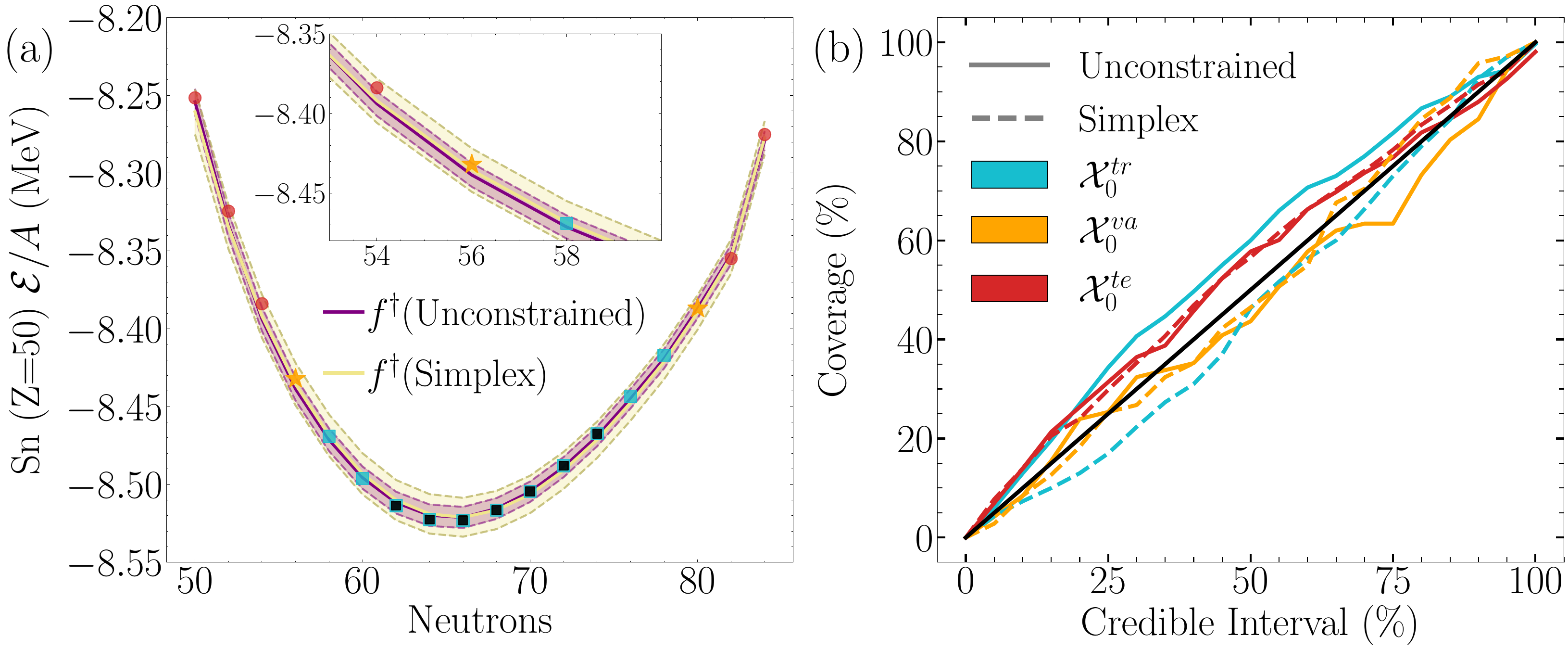}
		\caption{{\bf (a)}: Predictive posterior distribution for the binding energy per nucleon of the Sn isotopes. The mean prediction and $95\%$ credible interval of the unconstrained combined model is shown in purple, while  the simplex-constrained (simplex) combined model is shown  
 in khaki.
 The inset shows the detail of the plot  for $N=54, 56$, and 58.
 {\bf (b)} ECP for unconstrained and simplex-constrained variants  for training (blue), validation (yellow), and testing (red) datasets. The diagonal black line shows a reference of what a perfect statistical coverage would entail, with points above it being conservative, and those below being overconfident. 
  }
		\label{fig:bmmreal}
	\end{centering}
\end{figure*}

From the distributions of the weights
 \eqref{Eq: posterior}, we compute the predictions of the combined model $f^\dagger$ with quantified uncertainties across the entire dataset. Figure~\ref{fig:bmmreal}(a)  shows the predictions and $95\%$ credible intervals for the Sn ($Z=50$) isotopic chain for the unconstrained and 
 simplex-constrained variants. One feature to note is that 
 the simplex-constrained  model has a systematically broader credible interval in its forecasts both inside and outside the training region, yet both models seem to cover the experimental data well within their credible bands.
Since the constrained model can only reproduce a subset of possible combinations, we expect its performance in terms of RMSE to be less expressive in both interpolation and extrapolation, at the gain of less sensitivity to overfitting. Indeed, as can be seen in the last two rows of Table~\ref{tab: RMSE Realistic}, the RMSE scores for the constrained model are significantly worse than for the unconstrained
model - as well as the two best performing models FRDM-12 and HFB-24. Yet, they remain steady in the transition from training to validation and testing, while for the unconstrained approach the testing RMSE increases by about $20\%$ in comparison to the training RMSE.  That the unconstrained model $f^\dagger$ outperforms each individual model in terms of RMSE is not surprising -- an unconstrained combination of PCs fitted to a given dataset will outperform what each individual model can do alone, as has been shown for nuclear mass models in Ref.~\cite{Wu2024}.
Embedding the PCA-based model combination within a full Bayesian framework and the classification of data into training, validation, and test sets
then allows for more reliable forecasts with quantified uncertainties when extrapolating beyond experimentally known masses.

Indeed, by analyzing the empirical coverage of our calibrated model, we can quantitatively assess signs of overfitting. 
Figure~\ref{fig:bmmreal}(b)  shows the empirical coverage probability (ECP) ~\cite{Raftery2007,Gneiting2007} for both the unconstrained and simplex-constrained models across the three datasets considered. The fact that the empirical curves all lie close to the diagonal reference line gives us confidence that the combined predictions are neither being over confident (too small credible intervals) or over conservative (too big credible intervals). This is particularly reassuring for the test set, in which considerable extrapolations have been made (cf. Fig.~\ref{fig:4panel}(a)).
When considering just one data type (here: nuclear binding energies), the risk for overfitting is low,
yet if one wishes to consider model performance on quantities not in the original dataset, the risk for overfitting can be substantially increased.

\section{Conclusions and Outlook}
In this work, we propose, implement, and benchmark a Bayesian model combination framework accompanied by model orthogonalization using Principal Component Analysis. We discuss the features of the proposed BMC+PCA method by applying it to global models of nuclear binding energy.
Following the tests based on the analytic Liquid Drop Model, we carry out realistic BMC+PCA calculations of nuclear binding energies using  15 global computational nuclear models. We demonstrate that the BMC+PCA framework performs excellently in terms of prediction accuracy and uncertainty quantification.
While we have focused on the nuclear physics use case in this work, the method itself is completely general and can be applied broadly where model forecasts are utilized.
It is also easy to implement and results in an interpretable combination, simplifying the application to problems that span disciplines.
Furthermore, the computational scheme is robust against model repetitions  and it does not favor one model class when multiple copies are added. 
The  BMC+PCA technique can be reduced to BMM+PCA by imposing the simplex condition. In this case, only the several best performing models remain in the combination, and we recover an interpretation of the model combination that can be compared to other traditional BMA and BMM approaches where the \emph{positive} model weights are determined by the data. 
For this simplex constrained version both the Root Mean Squared Error (Table~\ref{tab: RMSE Realistic}) as well as its uncertainty bands (Figure~\ref{fig:bmmreal}) are bigger than the unconstrained version, yet the simplex constrained approach shows signs of less over-fitting in terms of extrapolation, with a performance that remains stable across the three sets (Panel (a) of Figure~\ref{fig:4panel}).

In addition to producing optimal multimodel forecasts with robust uncertainties, BMC+PCA is also capable of identifying model collinearities and redundancies, a functionality that stands to benefit other applications that aim to combine the wisdom of multiple nuclear models~\cite{Everett2021,Qiu2024,Cirigliano2022}. 
Furthermore, the framework is also able to perform model selection if the exact model happens to be present in the set of models.
The computational efficiency of the combined model also positions it well for wide distribution on web-based platforms, such as the Bayesian Mass Explorer project~\cite{bmex}.
While live evaluation of most of the individual models is impossible due to their inherent numerical complexity, the combined model can be evaluated, with uncertainties, on the fly.
The model combination procedure itself is also efficient, meaning interested users can provide their own datasets and update the resultant BMC+PCA model.

Future developments will include a local extension of BMC+PCA by assuming domain-dependent weights, i.e.,  $\PCParams_j \rightarrow \PCParams_j(\xb)$, see Ref.~\cite{Kejzlar2023}.
This enhancement will help construct model combinations that emphasize the local performance of models in certain regions, a typical scenario in physics modeling across multiple scales.
To aid adoption of the method, it is currently planned to implement the procedure into the open source model mixing software, \textsc{taweret}~\cite{Ingles:2023nha}.
We will also consider the extension to heterogeneous forecasts by considering data of several classes (e.g., other nuclear data like binding energies and charge radii).

\PRLsep

\begin{acknowledgements}
We thank  L\'eo Neufcourt for valuable suggestions during the early stages of the project. We also wish to thank the BMM working group of the BAND collaboration and Edgard Bonilla for useful discussions.
This material is based upon work supported by the U.S.
Department of Energy, Office of Science, Office of Nuclear
Physics under Awards Nos. DE-SC0023688  and DOE-DE-SC0013365, and
by the National Science Foundation under award number 2004601 (CSSI program, BAND collaboration).
\end{acknowledgements}

\bibliography{biblio}

\end{document}